\documentclass{article}
\pdfoutput=1

\usepackage{arxiv}

\usepackage[utf8]{inputenc} 
\usepackage[T1]{fontenc}    
\usepackage{hyperref}       
\usepackage{url}            
\usepackage{booktabs}       
\usepackage{amsfonts}       
\usepackage{nicefrac}       
\usepackage{microtype}      
\usepackage{xcolor}         
\usepackage{rotating}
\usepackage{subcaption}
\usepackage{multirow}
\usepackage{amsmath,amssymb}
\usepackage{enumitem}
\usepackage{caption}
\usepackage[stable]{footmisc}
\usepackage[ruled,vlined]{algorithm2e}
\usepackage{xspace}
\usepackage{amsbsy}
\usepackage{dsfont}

\newtheorem{thm}{Theorem}

\usepackage{graphicx}

\title{Random Offset Block Embedding Array (ROBE) for CriteoTB Benchmark MLPerf DLRM Model : 1000$\pmb{\times}$ Compression and 3.1$\pmb{\times}$ Faster Inference}

\newcommand{\rmafull}{Random Offset Block Embedding Array\xspace}
\newcommand{\rma}{ROBE\xspace}
\newcommand{\rmaa}{ROBE Array\xspace}
\newcommand{\rmaz}{ROBE-$Z$\xspace}
\newcommand{\hnet}{HashedNet\xspace}

\newcommand{\rmazbold}{ROBE-$\pmb{Z}$\xspace}

\newcommand{\ip}[2]{\langle #1,#2 \rangle}

\author{Aditya Desai \\
  Department of Computer Science\\
  Rice University\\
  Houston, Texas \\
  \texttt{apd10@rice.edu} \\
    \And
Li Chou\\
  Department of Computer Science\\
  Rice University\\
  Houston, Texas \\
  \texttt{lchou@rice.edu} \\
    \And
Anshumali Shrivastava\\
  Department of Computer Science\\
  Rice University\\
  Houston, Texas \\
  \texttt{anshumali@rice.edu} \\
}

\begin{document}
\maketitle
\begin{abstract}

Deep learning for recommendation data is one of the most pervasive and challenging AI workload in recent times. State-of-the-art recommendation models are one of the largest models matching the likes of GPT-3 and Switch Transformer. Challenges in deep learning recommendation models (DLRM) stem from learning dense embeddings for each of the categorical tokens. These embedding tables in industrial scale models can be as large as hundreds of terabytes. Such large models lead to a plethora of engineering challenges, not to mention prohibitive communication overheads, and slower training and inference times. Of these, slower inference time directly impacts user experience. Model compression for DLRM is gaining traction and the community has recently shown impressive compression results.  In this paper, we present \rmafull (\rma) as a low memory alternative to embedding tables which provide orders of magnitude reduction in memory usage while maintaining accuracy and boosting execution speed. \rma is a simple fundamental approach in improving both cache performance and the variance of randomized hashing, which could be of independent interest in itself. We demonstrate that we can successfully train DLRM models with same accuracy while using $1000 \times$ less memory. A $1000\times$ compressed model directly results in faster inference without any engineering effort. In particular, we show that we can train DLRM model using \rmaa of size 100MB on a single GPU to achieve AUC of 0.8025 or higher as required by official MLPerf CriteoTB benchmark DLRM model of 100GB while achieving about $3.1\times$ (209\%) improvement in inference throughput.

\end{abstract}
\twocolumn
\section{Introduction}
\vspace{-0.2cm}
Recommendation systems are one of the top applications of machine learning. For example, Facebook reports that recommendation inference accounts for over 79\% of AI inference cycles~\cite{gupta2020architectural}. Therefore, considerable efforts have been and continue to be expended to develop systems that help users make more personalized and well-informed choices in various application domains. Recent approaches utilize deep learning-based models to achieve state-of-the-art performance. However, a key challenge is the need to handle millions of categorical features that dominate the recommendation data~\cite{DLRM19, cheng2016wide}. Following the work in natural language processing~\cite{word2vec,transformer-17}, current approaches~\cite{DLRM19, DCN17, song2019autoint, guo2017deepfm, lian2018xdeepfm,huang2019fibinet} utilize a real-valued feature vector (i.e., embedding) to represent each categorical token. These categorical representations are learned, end-to-end, and organized in a memory structure called {\em embedding} tables.

\textbf{Production scale models have large storage cost.}
If the set of all categories is $S$ and the embedding dimension is $d$, then the embedding table size is $|S| {\times}d$. With the number of categorical tokens per feature as large as tens of millions, embedding tables consume over 99.9\% of total model memory. Specifically, memory footprint for models that utilize embedding tables can easily surpass hundreds of terabytes (TB)~\cite{mudigere2021high,DLRM19,MDTrick19,QuoRemTrick19, gupta2020architectural}. For example, Facebook recently showcased training of a 50TB sized model distributed over 128 GPUs~\cite{mudigere2021high}. 

\textbf{Inference with deep learning recommendation models is memory-bound.}
Access of embedding tables do not follow any recognizable pattern. Namely, the access is highly irregular.
The large size of embedding tables coupled with irregular and sparse access causes high cache miss rates~\cite{gupta2020architectural}. In fact, production scale systems spend 80\% of inference cycles in embedding lookups~\cite{gupta2020architectural}. Hence, these models are memory bound.

\textbf{Training deep learning recommendation models suffers from high communication cost.}
If we are using data parallel model training, then, at each step, we need to communicate the gradients of all the parameters to all the involved nodes. The communication cost is exactly equal to the size of the model. Therefore at each step of training industrial scale model, TBs of data need to be communicated. Moreover, the embedding tables for industrial scale models (multiple TBs) cannot be stored on a single GPU/node. Thus, the model has to be distributed on multiple nodes/GPUs and trained in a model parallel fashion as well. This adds additional communication cost to the forward and backward pass making training slow.

\textbf{Training of deep learning recommendation models is not accessible to a general user.}
Training models with large number of parameters, and on terabytes of data, comes with significant engineering challenges. In addition, such a task requires expensive hardware. Deep learning recommendation models have to be trained in a mixed model and data parallel setting on clusters of nodes or GPUs, which is cost prohibitive. Thus, these models are out of the reach for machine learning users without such access. This also severely restricts the possibility of fast research in this area.

The deep learning recommendation model (DLRM) architecture~\cite{DLRM19} gave rise to an increased interest in constructing more memory-efficient embeddings. Recent state-of-the-art efforts in this direction include increasing expressive power of embeddings by using additional computing over smaller memory such as  compositional embedding~\cite{QuoRemTrick19}; learning different sized embeddings for different values to leverage the inherent power law in frequencies ~\cite{MDTrick19,md1,md2,md3,md4,md5}, low rank decomposition of embedding tables ~\cite{ttrec}. These approaches show a single ($\approx{10}\times$, \cite{QuoRemTrick19, MDTrick19}) or double order ($\approx{100}\times$, \cite{ttrec}) of magnitude reduction in embedding table size with no (or minimal) loss of accuracy. In our empirical evaluation, we show that with \rmaa for DLRM model, we can obtain as much as $1000\times$ compression with similar (or even improved) accuracy, at the same time giving a multi-fold increase in the inference throughput performance. Specifically, we can train $1000\times$ compressed DLRM MLPerf model for CriteoTB dataset which reaches the same MLPerf AUC value 0.8025 or higher with a inference throughput boost of 3.1$\times$ . Also, similar observations can be made on Criteo Kaggle dataset where $1000\times$ compressed model can achieve similar or better accuracy as original model over variety of state of the art deep learning recommendation models.

\textbf{What are the implications of 1000$\pmb{\times}$ compression of embedding tables?}

\textbf{(1) Eliminate the need of model parallel training.} For models as large as 50TBs, a 1000$\times$ compression can reduce the model size to 50 gigabytes (GB), which can easily fit on a single high-end GPU (e.g., Nvidia A100). Hence, we can simply run a data-parallel model optimization. 

\textbf{(2) 1000$\boldsymbol\times$ lower communication cost.}
With data-parallel model optimization, we would achieve a 1000$\times$ reduction in communication cost at each step of model update. Therefore, this leads to significant savings in communication cost.

\textbf{(3) Lower memory latency.} In their paper, \cite{gupta2020architectural}, authors reveal that their production scale recommendation show cache miss rate of 8 MPKI (misses per 1000-instructions) as compared to 0.5 MPKI in RNN, 0.2 MPKI in FC and 0.06 MPKI in CNN. This high cache miss rate is the main cause of higher memory latency. With smaller memory footprint, we can potentially store the embedding tables or large parts of them on chip memory thus thwarting the problem at its root. In our experiments, we show a $3.1 \times$ speedup in inference with $1000 \times $ compression. 

\textbf{(4) Overall faster training and inference times.} 
Overall, we have the potential to construct compact models that have faster inference and training time. In our experiments, we will show $3.1\times$ improvement in inference throughput. Our proof-of-concept code does not show any improvement in training time per iteration. We leave optimizing training time for future work.

\textbf{(5) Faster refresh cycle for industrial models.}
With changing interests, recommendation data suffers from frequent concept shift \cite{conceptdrift}. Faster training implies we can now refresh models at a faster rate, and thus provide better service to the users of recommendation system.



\textbf{Our approach:} 
Weight sharing is a widely used idea in machine learning to reduce memory required for the model. Some examples include feature hashing \cite{featurehash} to reduce input dimension, \hnet \cite{hashtrick} to compress fully connected multi-layered neural networks, usage of filters in convolution neural networks, and recently demonstrated some success with LSH based weight sharing in recommendation models~\cite{desai2021semantically}. In this paper, we introduce a memory sharing technique -- \rmafull (\rma). We use universal hash functions on chunks/blocks of the embeddings in the embedding table to locate it in a small circular array of memory. We refer to this form of hashing as \rma hashing. In a standard feature hashing scenario, where we project a vector in higher dimension to lower dimension, \rma hashing outperforms the usual feature hashing as defined in \cite{featurehash}. We discuss the theoretical results in Section \ref{sec:theory}. In addition to being theoretically superior, \rma also leads to better cache performance due to coalesced array access ($3.1\times$ boost in inference throughput). Our results shed new light on how to make randomized hashing algorithms cache friendly, and at the same time, have superior variance. We also provide precise quantification of various trade-offs involved. The results could be of independent interest to algorithms community working on randomized hashing algorithms.


\textbf{Caveat: } The caveat while using our compression technique in recommendation models is that while training the model, we require more iterations than those required for training the original model. For example, the original CriteoTB MLPerf model (100GB) takes 1 epoch to reach the target AUC of 0.8025, while the same model using 1000$\times$ less memory with \rmaa (100MB) takes 2 epochs to reach the same AUC. We see similar trend in our experiments with the Criteo Kaggle dataset.  While we see a clear improvement of inference throughput ($3.1 \times$), our current proof-of-concept code does not show training time benefits. 

We believe that this 2x epochs might still be cheaper in terms of time, if we consider the post-processing efforts invested in model reduction techniques to make these models leaner. Even if we ignore post-processing costs, with memory optimizations leveraging on 1000$\times$ less memory footprint, we believe we should be able to get faster end-to-end training times even while requiring more number of iterations.  We leave this aspect for future work.
\begin{figure}
    \centering
    \caption{ \rma-D : Working of \rma-D (A) \textbf{Forward pass: } embedding for a token x is extracted from the location specified by the hash function, $h(x)$. The \rma array is circular, so the embeddings that overflow are continued from beginning.  (B) \textbf{Backward Pass: } the gradients of each of the embeddings are mapped back into the array and aggregated (via sum) into the \rma array.}
    \label{figure:rmad}
    \includegraphics[scale=0.18]{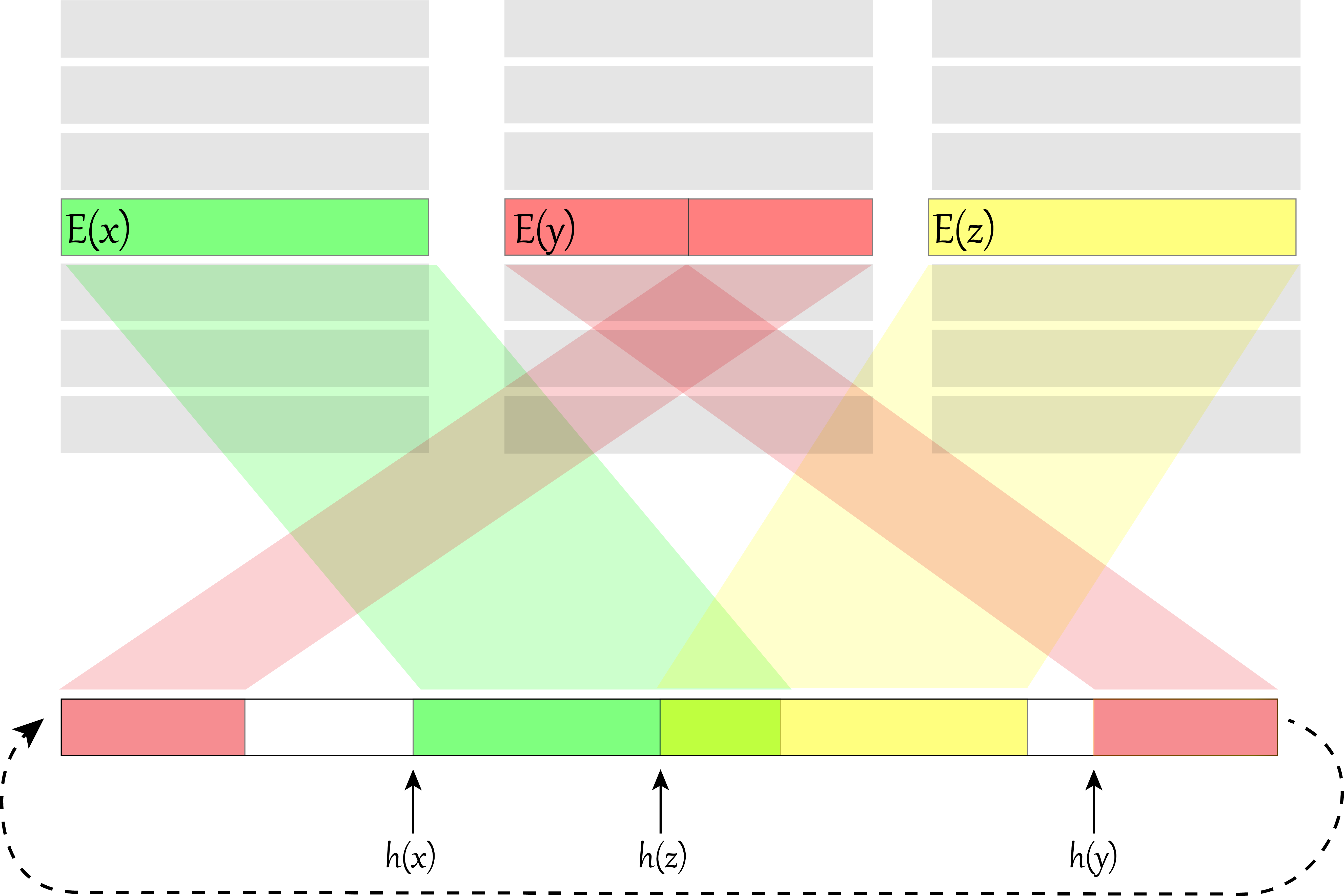}
    \vspace{-0.7cm}
\end{figure}
\vspace{-0.2cm}
\section{Related Work}
\vspace{-0.1cm}

The related lines of research in model compression can be broadly classified into two groups: (i) learning a compressed representation (or compression-aware training), and (ii) compressing the learned model via post processing. \rma-Z extends the former line of research, which has been widely applied to compressing DLRMs. Here, for models, we are referring to recommendation models. In addition, we focus our discussions on compressing embedding tables for these models.
\subsection{Learning compressed representations} 

\textbf{(1) Low rank decomposition:} Low rank decomposition of large matrices is a well known technique to reduce the memory footprint of the model. This entails representing the matrix under consideration, say $A \in \mathbb{R}^{D_1 \times D_2}$, as product of two low rank matrices, $B \in \mathbb{R}^{D_1 \times d}$ and $C \in \mathbb{R}^{d \times D_2}$ where $d \ll D_1, D_2$. In previous research on compressing DLRMs, low rank decomposition was applied in MD Embeddings \cite{MDTrick19} and TT-Rec \cite{ttrec}. In both works, embedding tables are first split by grouping together tokens based on their frequency in the dataset. Next, different rank-decompositions are applied to matrices representing parts of the embedding tables belonging to a particular group. The key idea is to use lower rank for tokens that appear sparsely in the dataset and hence leads to more memory saving. TT-Rec uses tensor train decomposition instead of the standard low rank decomposition to optimize for GPU computations. MD Embeddings has sparked a lot of related research for automatically partitioning the table and choosing the ranks of low-rank decomposition for optimal performance \cite{md1,md2,md3,md4,md5}. While MD Embeddings show compression of $16 \times$ without loss of quality, TT-Rec shows compression of $112 \times$ on Criteo Kaggle dataset. 

\textbf{(2) Feature hashing and compositional techniques :} An embedding table that is a tall matrix creates the problem of a high dimension input space. This problem has been traditionally solved in machine learning by feature hashing \cite{featurehash,shi09} where each input value is hashed to a smaller range using a hash function. This is quite similar to low rank decomposition where the first matrix is a fixed sparse matrix defined by the hash function. The authors of QR-Trick \cite{QuoRemTrick19} show that feature hashing does not work well for compressing embedding tables in recommendation systems. The reason is feature hashing forces the embedding of different tokens to be exactly same, thus causing a loss in the quality of model. QR-Trick ensures that each token gets a unique embedding by combining embeddings from multiple smaller embedding tables into a single embedding. The combining operation can be element-wise multiplication/ addition or even concatenation. QR-Trick gives $4\times$ compression with a slight loss in model quality.

\textbf{(3) \hnet} In their paper \cite{hashtrick}, authors introduced a technique to reduce memory usage of matrices in MLP networks. The weights of matrix are grouped randomly using a xxHash function and all the weights grouped together only use a single value from the underlying memory. This reduces the total memory foot print of the model. While this scheme is good to compress memory and reduce its footprint, it has some serious issues when it comes to efficiency. \hnet randomly distributes the elements of a matrix to varied locations. So, in order to access the a vector of size, say $d$, we have to potentially fetch d cache lines utilizing only $1/B$ fraction of bandwidth $B$. This large wastage of bandwidth can be one of the reasons why the community has not evaluated \hnet style compression with latency critical application such as recommendation. In this paper, we evaluate \hnet style compression and propose \rmaz scheme which is better than \hnet both in terms of quality and performance. Essentially, \rmaz achieve better quality of approximation than HashedNet while maintaining fraction of bandwidth usage of 1.

\textbf{(4) Quantization for training (low precision models)} : Research in reduced precision models for deep architectures has gained momentum recently \cite{lowprec1,lowprec2,lowprec3,lowprec4,lowprec5}. However, the challenges in recommendation models under consideration are unique and these techniques, developed primarily for CNN and RNN cannot naturally extend to recommendation. Recently, Facebook published their effort on using low precision models for DLRM in \cite{lowprec} and shows up to $2\times$ memory savings and $1.2 \times $ speed up.

\subsection{Compressing learned models}
These approaches require us to first train the baseline model and then compress them. Thus,  making these approaches less attractive to compression in recommendation models.

\textbf{(1) Quantization for inference:} Quantization can be performed post model training with the goal of reducing the inference time. The idea behind this quantization is to convert the floating point values to smaller representations, (e.g., int16 and int8), and replace floating point operations to integer operations, which are known to be faster. This approach can be applied in conjunction with the earlier approaches (learning compressed representation based) and \rma to improve performance further. 

\textbf{(2) Pruning:} Pruning \cite{pruning} compresses the model by removing edges from the computational graph of the model. It enables faster inference for models by reducing the computation. There is no straight forward way to apply pruning to compression of embedding tables and has not been explored in literature. 

\textbf{(3) Knowledge distillation:} Knowledge distillation as proposed by \cite{hinton2015distilling}, is a way of training a smaller model (called student) from a larger model (called teacher). Generally, the student model trained in this way outperforms the same model when trained standalone. One can imagine training a smaller dimensional embedding table from a larger embedding table. This approach has not been evaluated in the literature and can be explored further in an independent manner.

\vspace{-0.2cm}
\section{\rmafull}
\vspace{-0.1cm}

The memory footprint of the model is determined by the memory used to store the parameters of the model. In the case when the number of parameters far exceed the total amount of memory we intend to use, there are approaches such as mixed-precision learning \cite{lowprec}, low rank decomposition \cite{ttrec, MDTrick19} or specialized methods \cite{QuoRemTrick19} used to fit the parameters in the memory. To achieve order of magnitude more reduction in memory footprint of the model, we share memory among the elements of embeddings. Weight sharing scheme to compress MLP networks was proposed in \hnet \cite{hashtrick} but was never evaluated on embedding tables. We evaluate \hnet in our experiments and propose a weight sharing scheme that is provably better than standard weight sharing defined by \hnet, both in terms of quality and performance. 

Instead of storing an embedding table, we maintain a single array for learned parameters which is a compressed representation of embedding table. All embedding tables share the same array of learned parameters. The embeddings are accessed in a blocked manner from the embedding array using GPU-friendly universal hashing. We call this scheme of embedding compression as \rmafull (\rma). As we will see in Section \ref{sec:theory}, in learning a shared memory array via \rma , we can expect to get good quality models even with very high compression. This is further supported by our experiments in Section \ref{sec:experiments}.  What's more, \rma surmounts the memory bandwidth issues created by \hnet style hashing by making coalesced access.

The section is organized as follows. We first describe the most useful form of \rma i.e. \rma-D where D is the dimension of the embedding in embedding table. We then generalize the approach to consider \rma-Z for $Z \in \mathbf{N}$. In the next subsection, we contrast \rma-Z with \hnet style weight sharing. We end this section with a discussion on advantages of \rma-Z.

\subsection{\rma-D : \rma with block size equal to embedding size.}\label{sec:robed}

Consider that we are looking to build an embedding table of size $|S| \times D$ with embedding size D. We use a circular array $\mathcal{M}$ to store the learned parameters. Let $h : \mathbf{N} \rightarrow \{0,..,m-1\}$ be a hash function drawn uniformly randomly from a universal hash family. Similarly let $g: \mathbf{N \times N} \rightarrow \{-1,1\}$ be an independent hash function drawn from a different hash family with range 2. The working of \rma-D is illustrated in the figure \ref{figure:rmad}. 

\textbf{Forward Pass:} The embedding for a given token x as a whole is located in the \rma array using a universal hashing function, $h$. In case, the $h(x) + D \geq |\mathcal{M}|$, the embedding is continued in the first part of the \rma array. The embedding can, optionally, be multiplied element-wise with a value from $\{+1, -1\}$ as obtained via the hash function $g(x,i)$. Let vector $\mathbf{G}(x) = \{g(x, 1), g(x, 2) , ... , g(x, D)\}$.  If P(x) is a primary embedding obtained from the \rma array. Then the final embedding can be considered as $\mathbf{E}(x) = \mathbf{G}(x) \circ \mathbf{P}(x) $ where $\circ$ is element wise multiplication. Thus, we can write
\begin{equation}
\begin{split}
    &\mathbf{P}(x) = \mathcal{M}[h(x):h(x)+D]  \quad \mathrm{if} \quad h(x) + D < |\mathcal{M}| \\
    & \mathbf{P}(x) = p_1.p_2 \quad \mathrm{if} \quad h(x) + D \geq |\mathcal{M}| \\
    &\quad \quad \quad where \\
    &\quad \quad \quad \quad p_1 = \mathcal{M}[h(x):|\mathcal{M}|] \\
    &\quad \quad \quad \quad p_2 = \mathcal{M}[0:(D - (|\mathcal{M}| - h(x)) )] \quad \\
    & \mathbf{E}(x) = \mathbf{G}(x) \circ \mathbf{P}(x)
\end{split}
\end{equation}
where "." denotes concatenation.

\textbf{Multiple embedding tables:} As discussed earlier, all embedding tables in the system, even with varying embedding dimensions, will share the \rma array. In order to achieve independent locations for embeddings in each table, the hash function $h$ and $g$ is modified to include embedding table id as a parameter $h: N \times N \rightarrow \{0, ..., m-1\}$, $g: N \times N \times N \rightarrow \{+1, -1\}$

\textbf{Hashing functions $\mathbf{h}$, $\mathbf{g}$ and $\mathbf{\mathrm{idx}}$ mapping}. We use universal hash function families to choose $h$ and $g$. The significant aspect of universal hash functions is that they are cheap to compute, GPU-implementation friendly and provide decent guarantee-bound on collision probability. With universal hash functions for $h$ and $g$ the mapping, $\mathrm{idx}$, for element $i$ of the embedding for a token $x$ from embedding table $e$ can be written as,
\begin{equation}
    \begin{split}
        &h(e, x) = ((A_h e  + B_h x  + C_h) \mod P \mod |\mathcal{M}| \\
        &\mathrm{idx}(e,x,i) = (h(e, x) + i)  \mod |\mathcal{M}| \\
        &g(e, x, i) = 2 ( \\
        & \quad \quad \quad ((A_g e  + B_g x  + C_g i + D_g) \mod P \mod 2\\
        & \quad \quad \quad ) - 1 
        \end{split}\label{eq:eq_rmax}
\end{equation}
where P is a large prime and $A_h, B_h \in \{1, ... P-1\}, C_h \in \{0, ..., P-1\}$ are randomly chosen values. Thus, $i^{th}$ element of the embedding of x from embedding table e can be written as $\mathcal{M}[\mathrm{idx}(e, x, i)] * g(x, i)$

\textbf{Backward Pass:}
The backward pass can also be illustrated with the same figure \ref{figure:rmad}. Essentially, the gradients of the embeddings are mapped into the \rma array according to the $\mathrm{idx}$ and undergo a signed aggregation.
\begin{equation}
    \Delta(\mathcal{M}[j]) = \sum_{\mathrm{idx}(e,x,i) = j} g(e, x, i) \Delta(\mathbf{E}_e(x)[i]) \label{eq:backward}
\end{equation}
where $\mathbf{E}_e$ represents the embedding table with table id $e$ and $\Delta(p)$ is the gradient of the loss function w.r.t the parameter p.
\begin{figure}
    \centering
        \caption{ \rma-Z : Working of \rma-Z (A) \textbf{Forward pass: } embedding for a token x is assembled by extracting chunks of memory from the locations specified by the hash function, $h(x,c_{id})$. The \rma array is circular, so the embeddings that overflow are continued from beginning.  (B) \textbf{Backward Pass: } the gradients of each of the embeddings are mapped back into the array and aggregated (via sum) into the \rma array.}
    \label{figure:rmaz}
    \includegraphics[scale=0.18]{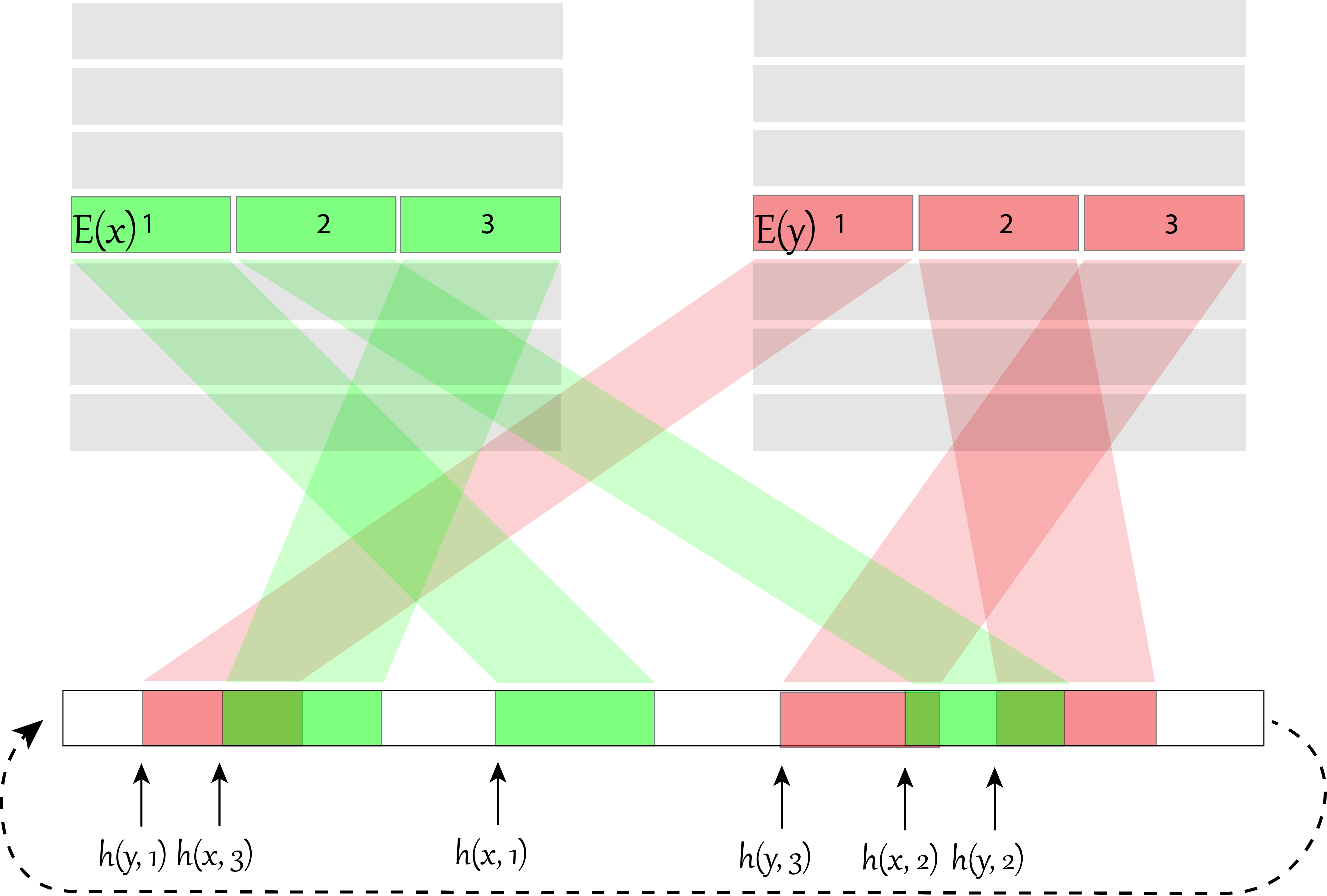}
    \vspace{-0.8cm}
\end{figure}
\subsection{\rma-Z : \rma with arbitrary block size $Z \leq D$} \label{sec:robez}
    \begin{table*}[]
    \centering
    \caption{Number of memory fetches on varying sizes of $Z$, where $D$ is the embedding size, $B$ is the bus size and $Z$ is the block size. $B | D$ denotes $B$ divides $D$.}
\begin{tabular}{|l|l|l|l|}
\hline
                          & \textbf{Condition} & \textbf{Max number of memory fetches} & \textbf{Comment}                                          \\ \hline
Original                  & $B | D$              & $D/B + 1$                               &                                                           \\ \hline
\hnet/ROBE-1         & &  $D$                               & \\ \hline
\rmaz $Z < B < D$ & $Z | B | D$          & $2 \times D/Z$                               & With high probability as $|M| \gg d >  Z$ \\ \hline
\rmaz $B < Z < D$ & $B | Z | D$          & $D/B + D/Z$                           & With high probability as $|M| \gg d > Z$ \\ \hline
\rmaz $Z \geq D$                & $D | Z$              & $D/B + 2$                             &                                                           \\ \hline
\end{tabular}
\label{tab:memfetches}
\end{table*}
We generalize \rma-D to \rma-Z with arbitrary block size. In this case, the embedding of a particular token is split into chunks of size Z. These chunks are independently mapped into the \rma array using universal hash functions similar to \rma-D. The procedure is illustrated in figure \ref{figure:rmaz}. Borrowing notation from the previous section, 
\begin{equation}
\begin{split}
    &\mathbf{P}_{c}(x) = \mathcal{M}[h(x, c):h(x, c) + Z]  \textrm{  } \mathrm{if} \textrm{ } h(x,c) + Z < |\mathcal{M}|\\
    & \mathbf{P}_{c}(x) = p_1.p_2 \quad \mathrm{if} \quad h(x) + Z \geq |\mathcal{M}| \\
    &\quad \quad \quad where \\
    &\quad \quad \quad \quad p_1 = \mathcal{M}[h(x):|\mathcal{M}|] \\
    &\quad \quad \quad \quad p_2 = \mathcal{M}[0:(Z + h(x) - |\mathcal{M}| )] \quad \\
    & \mathbf{P}(x) = \mathrm{P}_{c_1}.\mathrm{P}_{c_2}.\mathrm{P}_{c_3}...\mathrm{P}_{c_{D/Z}}\\
    & \mathbf{E}(x) = \mathbf{G}(x) \circ \mathbf{P}(x)
\end{split}
\end{equation}
where "." denotes concatenation.
The $\mathrm{idx}$ mapping for a multiple table \rma-Z can be written as, 
\begin{equation}
    \begin{split}
        & \mathcal{Z}_{id}(i) = \lfloor i/Z \rfloor \quad \mathcal{O}_{id}(i) =  i \mod Z  \\
        & h(e, x, c) = ((A_h e  + B_h x  + C_h c + D) \mod P \mod |\mathcal{M}| \\
        &\mathrm{idx}(e,x,i) = (h(e, x, \mathcal{Z}_{id}(i) + \mathcal{O}_{id}(i)))  \mod |\mathcal{M}| 
        \end{split}\label{eq:eq_rmax}
\end{equation}
where P is a large prime and $A_h, B_h, C_h \in \{1, ... P{-}1\}, D_h \in \{0, ..., P{-}1\}$ are randomly chosen values. In the equations above, $\mathcal{Z}_{id}(i)$ represents the chunk id of the index $i$ and $\mathcal{O}_{id}(i)$ computes the offset within the chunk of i.

The expression for g(x, e, i) is same as in \rma-D. Also, the backward pass functions similarly to \rma-D according to equation \ref{eq:backward}

\subsection{\rma-1 vs \hnet}
\rma-1 hashing scheme is similar to the hashing scheme proposed by \hnet \cite{hashtrick} for compressing matrices in MLP network. There are some differences though. \rma-1 uses light weight universal hashing as opposed to xxHash used by \hnet. Thus, \rma-1 compromises the collision guarantees for better performance. 
Using universal hashing makes implementing the computation on GPU very convenient and efficient. Additionally, \hnet, demonstrated on MLP networks,  keeps separate arrays for separate matrices, whereas \rma-1 use a single array to map all the elements from all the embedding tables. What is most exciting about \rma approach is its \rma-Z ( or \rma-D ) which is theoretically superior to hashing proposed by \hnet and is cache efficient due to appropriate usage of cache-lines via coalesced access.

The setup can also be extended to $Z>D$ by clubbing multiple embeddings together in a chunk. The formulation follows the same scheme as shown in sections \ref{sec:robed} and \ref{sec:robez} This actually leads to better feature hashing quality as shown in section \ref{sec:theory}.

\subsection{Advantages of \rmazbold}

\textbf{Memory Latency and Issue of Irregular memory access}
As mentioned in \cite{gupta2020architectural}, recommendation models suffer a very high cache-miss rate due to large embedding tables and irregular access as compared to other architectures. \rmaz can partially solve this problem by potentially storing large part of the embedding table (or even entire embedding table) in a compressed format in LLC . For example, embedding tables with a collective size of 100GB, when allocated a memory of 100MB (i.e. $1000\times$ reduction), can be stored on last level cache. The original model, in this case, without any memory sharing has to be stored on RAM, or even worse on disk.

\textbf{Better compute intensity} In their paper, \cite{gupta2020architectural}, authors highlight low compute intensity as one of the unique challenges in embedding tables. With reusing a lot of memory locations, \rmaz improves the compute intensity of the embedding tables.

\textbf{Memory Fetches:}
The number of memory fetches can potentially increase when using a \rmaz allocation scheme, especially worse during \rma-1 (or \hnet). The reason is wasting band-width of cache line. We present the number of cache-line fetches while using \rmaz and compare it against the memory fetches with using original embedding and \hnet, which is shown in Table \ref{tab:memfetches}.

Consider the original embedding of size $D$ (generally kept in multiples of cache-line size). Let the cache-line size be $B$. Thus, in order to fetch a single embedding from original embedding table, we would require a maximum of $(D/B + 1)$ ( +1 for non-aligned access) memory fetches. As we can see from Table \ref{tab:memfetches}, as we increase the value of $Z$ the number of cache-line fetches decrease from the $2D/Z$ to $D/B+2$ due to the coalesced access pattern when $Z$ is greater than $D$. Also, as we will see in section \ref{sec:theory}, the greater the value of $Z$, the better is dimensionality reduction. So it is advisable to choose a large value for $Z$.

\textbf{Dimensionality Reduction:} As we will see in Section \ref{sec:theory}, \rmaz hashing is better than \rma-$1$ in terms of dimensionality reduction. As the value of $Z$ increases, while the estimate of inner products in projected space is unbiased, the variance decreases until $Z$ reaches $|M|$.

\vspace{-0.2cm}
\section{Theoretical Considerations}\label{sec:theory}
\vspace{-0.1cm}

\begin{figure}
        \centering
        \caption{Sketching matrix without sign for \hnet /feature hashing/ \rma-1 (left) and \rma-3 (right).}
        \includegraphics[scale=0.3]{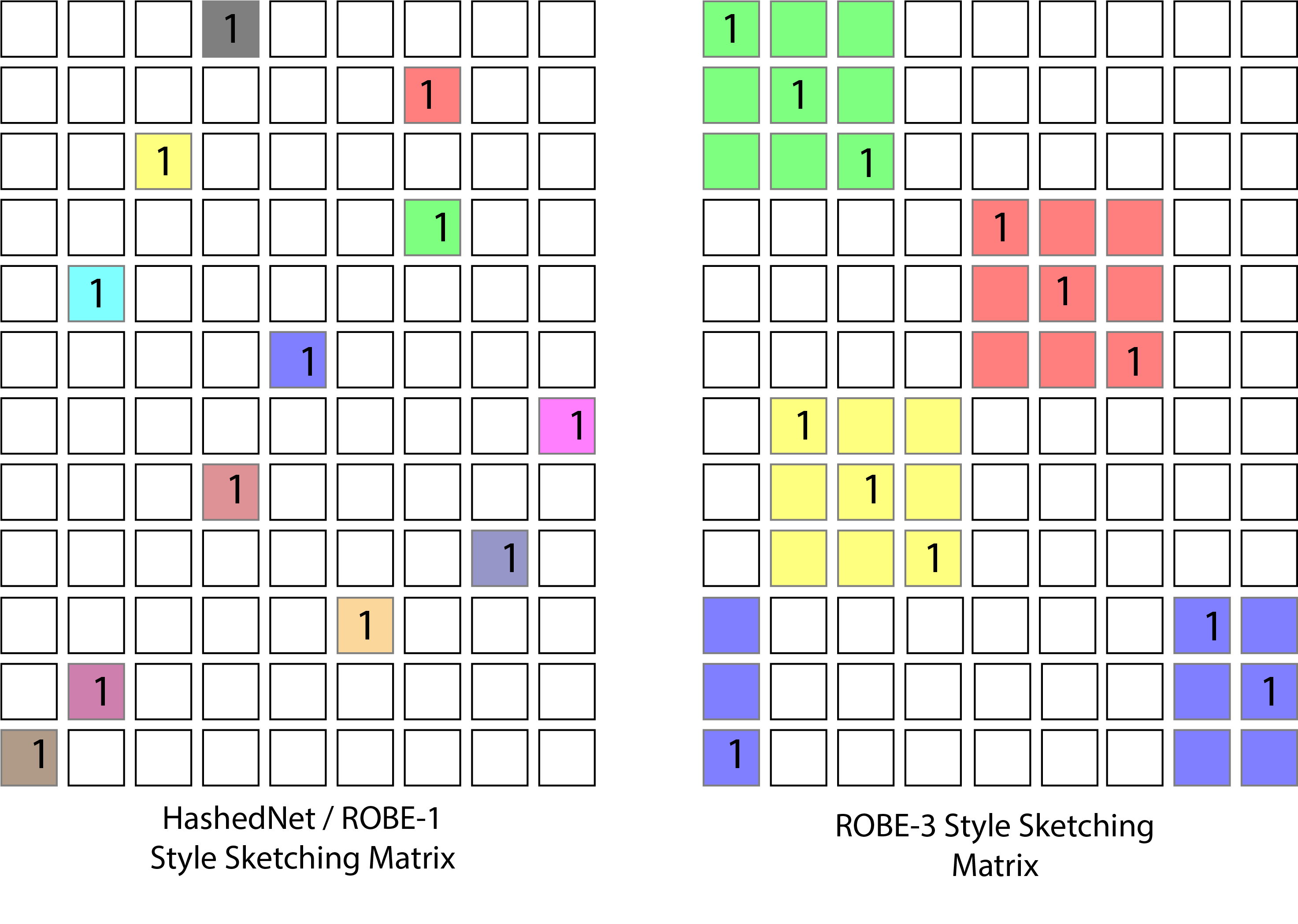}
\label{fig:sketchmatrix}
\vspace{-0.8cm}
\end{figure}

    The procedure described in \rmaz is closely related to the sketching literature, and in particular, the area of random projections. A \textit{parameter vector} can be created by joining all the flattened embedding matrices. The \rmaz hashing, essentially, projects this parameter vector into a $\mathbb{R}^{|\mathcal{M}|}$ space. We know from Johnson-Lindenstrauss Lemma, that random projections can provide us with low-dimensional and low-distortion embeddings of vectors from high dimension. Feature hashing is an efficient form of random projection where the sketching matrix is sparse - i.e. each row of the matrix has exactly one non-zero (usually $\pm 1$) and this location is determined randomly. We can visualize the sketching matrix for \rmaz as shown in Figure~\ref{fig:sketchmatrix}
.
    Using the mapping function, or alternatively the sketching matrix, we can recover the original embedding vector from the memory. In fact, in this paper, we directly learn the compressed representation of the parameter vector (hence embedding tables).
    
    We provide two analysis. (1) The first analysis measures the quality of the dimensionality reduction while using \rmaz hashing. This is a standard analysis on the lines of that presented in \cite{featurehash}. We show that \rmaz ( with $Z > 1$) is better than \rma-1 which is essentially the feature hashing described in \cite{featurehash}. 
    
    (2) The previous papers such as \hnet only evaluate their method on dimensionality reduction. However, depending on how the compressed memory is being used, we believe it is important to also measure the application specific effect of compression. Hence, we analyse the quality of embedding structure maintained by the \rmaz hashing in the projected space. In this analysis, we measure how the relation between two embeddings is maintained under this memory allocation scheme.
    
    \subsection{Dimensionality reduction : \rmazbold beats feature hashing}
    In order to assess the quality of dimensionality reduction of the parameter vector in $\mathbb{R}^n$, we look at the estimation of the inner product of two vectors in the projected space. This is a standard way to measure the preservation of distances under projection.
    
    Let $x$ and $y$ be two parameter vectors in $\mathbb{R}^n$. Note that these are not embedding vectors but two parameter vectors. Let the inner product between $x$ and $y$ be denoted as $\ip{x}{y}$. Let \rmaz sketching matrix be a $n \times m$ matrix projecting the vectors in space of $\mathbb{R}^n$ to $\mathbb{R}^m$ such that $m < n$. Let the projected vectors be $\hat{x}$ and $\hat{y}$ respectively. The projections are obtained by using two hash functions $h$ and $g$. $h$ is the memory allocation function described in section above, which maps the block id into the range $\{0,...,m-1\}$, and $g$ assigns a value in $\{+1,-1\}$ to each index of the vector. The final mapping obtained is $\mathrm{idx}$ as described in section \ref{sec:robez}. Note that while hash function $h$ applies to blocks, $g$ gives independent values to each index within the block as well. We do not use $g$ in practice, but here we use $g$ to simplify the analysis. To begin with, the inner product estimator can be written as
\begin{equation*}
\begin{split}
    \widehat{\ip{x}{y}} = \sum_{j = 1 }^m & \left( \sum_{i=1}^n x_i g(i)  \mathds{1}(\mathrm{idx}(i) == j) \right) \\
     & \quad \quad \left( \sum_{i=1}^n y_i g(i) \mathds{1}(\mathrm{idx}(i) == j) \right)
    \end{split}
\end{equation*}
    where $\mathds{1}(\cdot)$ is an indicator function. The expectation and variance of the inner product estimator can be written as below (Theorem~\ref{thm:exp_var}). Here, we use the function $\mathcal{Z}_{id}$ directly on the index of parameter vector as opposed to how it is defined in Equation (\ref{eq:eq_rmax}) when we are dealing with embedding tables. Both notations are equivalent and we use them interchangeably depending on context.
    
    \begin{thm}
    \label{thm:exp_var}
    The inner product of two parameter vectors $x, y \in \mathbb{R}^n$ projected into the space $\mathbb{R}^m$ using the sketching matrix for \rmaz with block size $Z$, $m < n$ and $Z < m$ is a random variable with expectation and variance as noted below. Let $\mathcal{Z}_{id}(i)$ denote the block id of index $i$ as computed in Equation (\ref{eq:eq_rmax}).
    \begin{align*}
    &\mathbb{E}(\widehat{\ip{x}{y}}) = \ip{x}{y}\\
    &\mathbb{V}(\widehat{\ip{x}{y}}) = \frac{1}{m} (\Sigma_{\mathcal{Z}_{id}(i) \neq \mathcal{Z}_{id}(j)}  x_i^2 y_j^2 +  \Sigma_{\mathcal{Z}_{id}(i) \neq \mathcal{Z}_{id}(j)} x_i y_i x_j y_j)
\end{align*}
Let $\mathbb{V}_Z(x,y,n,m)$ be the variance of inner product between $x, y \in \mathbb{R}^n$ while using \rma-Z on memory size $|M|=m$. Then the above equation can be rewritten as follows. 
\begin{equation*}
\begin{split}
  \mathbb{V}_Z&(x, y, n, m) = \mathbb{V}_1(x,y,n,m) - \\ 
 & \sum_{i=0}^{n/Z-1} \mathbb{V}_1(x[Zi:Z(i+1)],y[Zi:Z(i+1)],Z,m),
 \end{split}
\end{equation*}
where $x[i:j]$ refers to slice of vector from index $i$ to $j$.
\end{thm}


Note that $\mathbb{V}_1(x,y,n,m)$ is exactly the variance observed with \rma-1 or feature hashing matrix \cite{featurehash}. As we can clearly see that \rmaz has lower variance than \rma-1. Hence, \rmaz is better at dimensionality reduction than feature hashing.

\textbf{Intuition:} The results are not surprising and can explained by observing that once we hash blocks, we ensure that elements of parameter vector that lie within a particular block do not collide under random projection (benign correlations due to blocking). Also, the marginal probability of collision of two elements that lie in different blocks is same as that in \rma-1. While there is additional constraint in \rmaz of the form ``if $i$ and $j$ collide then $i+1$ and $j+1$ also collide if they lie in same blocks as $i$ and $j$ respectively,'' these relations do not affect the variance as can be seen in detailed analysis in the Appendix. The exciting part is that the improved variance also comes with improved cache performance.

The analysis can be extended to get concentration inequalities on the lines of the analysis provided by \cite{featurehash}. 
\subsection{Embedding structure preservation under projection}
Let $a$ and $b$ be two tokens and their corresponding embedding vectors be $\vec{\theta_a}, \vec{\theta_b} \in \mathbb{R}^D$ . Let $\theta$ be the parameter vector, then $\theta_a = \theta[id_a:id_a+D]$ for some $id_a$. We assess the quality of embedding structure preservation under the \rmaz projection by measuring the inner product estimation between $\theta_a$ and $\theta_b$ under the projection. The inner product estimator can be written as
\begin{equation*}
\begin{split}
     &\widehat{\ip{\vec{\theta_a}}{\vec{\theta_b}}} =  \\
     &\sum_{j=1}^D \left( \sum_{i=1}^n \theta_i g(id_a + j) g(i) \mathds{1}(\mathrm{idx}(i) == \mathrm{idx}(id_a + j)) \right) \\ 
    & \quad \left( \sum_{i=1}^n \theta_i g(id_b  + j) g(i) \mathds{1}(\mathrm{idx}(i) == \mathrm{idx}(id_b + j)) \right).
    \end{split}
\end{equation*}
\begin{thm}\label{thm:thm_two}
The inner product for embeddings of two distinct values $a$ and $b$, say $\theta_a, \theta_b \in \mathbb{R}^D$, when the parameter vector $\theta \in \mathbb{R}^n$ is projected onto a space $R^m$ using the \rmaz hashing has an expected value as shown below.
\begin{equation*}
    \mathbb{E}(\widehat{\ip{\vec{\theta_a}}{\vec{\theta_b}}}) = \ip{\theta_a}{\theta_b},
\end{equation*}
when $a$ and $b$ have embeddings in same block.
\begin{equation*}
    \mathbb{E}(\widehat{\ip{\vec{\theta_a}}{\vec{\theta_b}}}) = \ip{\theta_a}{\theta_b} (1 + \frac{1}{m}), \quad \quad 
\end{equation*}
when $a$ and $b$ have embeddings in different blocks or $Z < D$.
\begin{table*}[]
\caption{Comparison of ROBE-Z and other baselines on Criteo datasets.}
\label{tab:baselines}
\begin{tabular}{|l|l|l|l|l|}
\hline
Method        & Dataset       & Memory Compression & Quality of model              & Metric Used      \\ \hline
Low-Precision\cite{lowprec}      & Criteo Kaggle       & $2 \times $ &  similar to baseline    & Logloss      \\ \hline

Hashing Trick \cite{featurehash} & Criteo Kaggle & $4 \times$         & Much worse than QR Trick      & Logloss          \\ \hline
QR Trick   \cite{QuoRemTrick19}    & Criteo Kaggle & $4 \times$         & Slightly worse than baseline     & Logloss          \\ \hline
MD Emeddings \cite{MDTrick19}  & Criteo Kaggle & $16 \times$        & Better or Similar to baseline           & Logloss          \\ \hline
TT-Rec      \cite{ttrec}  & Criteo Kaggle & $112 \times$       & Better or similar to baseline & Accuracy/logloss \\ \hline
TT-Rec        & Criteo TB     & $117 \times$       & Better or similar to baseline & Accuracy/logloss \\ \hline
ROBE-Z        & Criteo Kaggle & $1000 \times$      & Better or similar to baseline & AUC/logloss      \\ \hline
ROBE-Z        & Criteo TB     & $1000 \times$      & Better or similar to baseline & AUC/logloss      \\ \hline
\end{tabular}
\end{table*}
The variance for the case when embeddings of $a$ and $b$ lie in separate blocks and $Z \geq D$ can be expressed as 
\begin{equation*}
\begin{split}
     \mathbb{V}({\ip{\vec{\theta_a}}{\vec{\theta_b}}})  \leq  \mathbb{O}\left( \frac{D}{m^2} ||\theta||^4 + \frac{1}{m^2} (\sum_{i\neq j = 1}^D \theta_a^{(i)} \theta_b^{(i)} \theta_a^{(j)} \theta_b^{(j)}) \right)
\end{split}
\end{equation*}
\end{thm}
We provide variance of inner product estimator for a commonly occuring case. Detailed proof of Theorem~\ref{thm:thm_two} can be found in the Appendix. Variance for other cases can be computed similarly. We factor out the dependence on Z to simplify the expression. However , it can be noted that if $Z < D$ there will be additional interactions which will increase the variance.

\textbf{Intuition:}
The expected value of inner product of two embeddings is unbiased only in the case when embeddings of $a$ and $b$ lie in the same block. This is expected as when embeddings of $a$ and $b$ lie in different blocks, $i^{th}$ element of $a$ and $b$ can be potentially mapped to the same location in memory leading to biased estimates. Also, the variance depends on the $\ell$2-norm of the parameter vector (equivalently frobenius norm of embedding tables). Again, one can expect this as every element can potentially collide with the elements of embeddings of $a$ and $b$. It is interesting to note that while dimensionality reduction's variance was proportional to $\frac{1}{m} ||\theta||^2$, the variance in this case is proportional to $(\frac{1}{m} ||\theta||^2)^2$. One can expect this to happen as in this analysis, vectors share the same memory ( as opposed to dimensionality reduction).

\begin{table*}[h]
\centering
\caption{Criteo Kaggle dataset: Test AUC of $1000\times$ reduced \rmaz model (2MB embedding) compared against original model (2GB). Epochs for \rmaz models measure epochs taken to reach AUC of original models. In case models do not reach quality of original model, we note the epochs to reach their best AUC. The AUC under \rmaz shows the best AUC reached (using early stopping). While all models overfit after a time, DLRM models do not and the training is cutoff at 11 epochs. All values of Z in \rmaz show similar std-deviation. So we only write one of them.} \label{tab:main}
\resizebox{\textwidth}{!}{
\begin{tabular}{|c|c|c|c|c|l|c|c|c|c|}
\hline
Model   & \begin{tabular}[c]{@{}c@{}}Original\\ \#1\end{tabular} & Epochs & \begin{tabular}[c]{@{}c@{}}\rma-1 \\ (avg)\\  (seed 1,2,3)\end{tabular} & \begin{tabular}[c]{@{}c@{}}\rma-1\\ (stdev)\end{tabular} & \begin{tabular}[c]{@{}l@{}}Epochs\\ (reach \#1)\end{tabular} & \begin{tabular}[c]{@{}c@{}}\rma-2 \\ (seed 1)\end{tabular} & \begin{tabular}[c]{@{}c@{}}\rma-4\\ (seed 1)\end{tabular} & \begin{tabular}[c]{@{}c@{}}\rma-8 \\  (seed 1)\end{tabular} & \begin{tabular}[c]{@{}c@{}}\rma-16\\ (seed 1)\end{tabular} \\ \hline
DLRM    & 0.8031                                                 & 1.37   & 0.8050                                                                                 & 0.0001                                                                  & 3.96                                                         & \textbf{0.8050}                                                           & 0.8049                                                                   & 0.8047                                                                     & \textbf{0.8050}                                                           \\ \hline
DCN     & 0.7973                                                 & 1      & 0.7991                                                                                 & 0.0004                                                                  & 1.8                                                          & 0.7994                                                                    & \textbf{0.7995}                                                          & 0.7994                                                                     & 0.7993                                                                    \\ \hline
AutoInt & 0.7968                                                 & 1      & 0.7987                                                                                 & 0.0002                                                                  & 1.62                                                         & 0.7984                                                                    & 0.7984                                                                   & \textbf{0.7988}                                                            & 0.7985                                                                    \\ \hline
DeepFM  & \textbf{0.7957}                                        & 1      & 0.7951                                                                                 & 0.0001                                                                  & 1.99                                                         & 0.7949                                                                    & 0.7949                                                                   & 0.7947                                                                     & 0.795                                                                     \\ \hline
XDeepFM & \textbf{0.8007}                                        & 1.625  & 0.7989                                                                                 & 0.0004                                                                  & 3.93                                                         & 0.7987                                                                    & 0.7988                                                                   & 0.799                                                                      & 0.7991                                                                    \\ \hline
FiBiNET & \textbf{0.8016}                                        & 3      & 0.8011                                                                                 & 0.0002                                                                  & 2.99                                                         & 0.8011                                                                    & 0.8010                                                                   & 0.8013                                                                     & 0.8012                                                                    \\ \hline
\end{tabular}}
\vspace{-0.5cm}

\end{table*}
\vspace{-0.2cm}
\section{Experimental Results}\label{sec:experiments}
\vspace{-0.1cm}

We evaluate \rmaz for embedding tables in deep learning recommendation models in this section. Subsection \ref{sec:baselines} compiles various baselines from literature which try to compress embedding tables for various recommendation datasets. One can clearly see that \rmaz provides the best compression of embedding tables that is orders of magnitude better than previous state of the art. In the section \ref{sec:tbacc} we show that quality of model holds for different values of Z on CriteoTB dataset for Facebook DLRM model. In the section \ref{sec:kagacc}, we show that the results of \rmaz holds on Criteo Kaggle dataset for a set of state-of-art recommendation models proving the generalized success of \rmaz. In the last two subsections \ref{sec:inf} and \ref{sec:train}, we discuss the effect of \rmaz on inference and training times respectively. Specifically, we show that using \rmaz gives $3.1 \times$ more throughput during inference.  The two datasets are described in sections \ref{sec:tbacc} and \ref{sec:kagacc}.

\subsection{Comparison with baselines} \label{sec:baselines}

We select the following baselines to compare our results against. More details on these methods can be found in the related work section of the paper. The results are aggregated in the table \ref{tab:baselines}
\begin{itemize}[nosep, leftmargin=*]
    \item \textbf{Quantization with low precision models:} In \cite{lowprec}, authors note that low precision embedding tables gives upto $2\times$ reduction in embedding tables by using FP16 instead of FP32.
    \item \textbf{ Hashing Trick} : This is the popular form of input size reduction technique as introduced in \cite{featurehash}. In this technique, each category value is hashed to a smaller range and all values that are mapped to a single value use the same embedding. As reported by the \cite{QuoRemTrick19}, hashing trick performs quite worse than baseline with a compression of $4\times$
    \item \textbf{Compositional Embeddings (QR Trick)} : This technique ensures that each value has a unique embedding by composing different chunks from shared pool of embeddings. As reported in \cite{QuoRemTrick19}, at $4 \times $ compression, we see slight drop in quality of the model as compared to original on Criteo Kaggle dataset.
    \item\textbf{ Mixed Dimensional (MD) Embeddings} :  We choose MD Embeddings \cite{MDTrick19} as a representative technique of different techniques \cite{md1,md2,md3,md4,md5} to choose mixed dimensional embeddings for compression. The paper reports $16 \times$ compression with results similar to the baseline on criteo Kaggle dataset.
    \item\textbf{ TT-Rec} \cite{ttrec} employs a tensor-train decomposition of the embedding matrix. As reported in \cite{ttrec}, it shows little over $100 \times$ compression on both Criteo Kaggle and Criteo TB datasets. This is an impressive improvement in memory usage as compared to its predecessors.
\end{itemize}

As can be seen in table \ref{tab:baselines}, ROBE-Z provides orders of magnitude improvement in memory compression while maintaining the quality of the model. Thus \rmaz beats the state of the art compression results on DLRM models.

\subsection{1000$\boldsymbol\times$ Compression of CriteoTB MLPerf Model with AUC 0.825 or higher with varying values of Z} \label{sec:tbacc}

\textbf{Dataset:} CriteoTB dataset has 13 integer features and 26 categorical features with around 800 million categorical tokens in total. This is the advertising data of 23 days published by criteo. We use exactly same setting as mentioned for official version by MLPerf for training.

\textbf{Model:} The official MLPerf model for DLRM on CriteoTB (see section \ref{sec:reproduce} in appendix for code details) requires around \textbf{100GB} sized embedding tables and achieves the target MLPerf AUC of 0.8025 in 1 epoch. We will use the same quality metric of 0.8025 AUC as prescribed in MLPerf settings for CriteoTB dataset to evaluate \rma-Z.

\textbf{Results:} With \rma-Z using 1000$\times$ less memory, i.e. \textbf{only 100MB}, we achieve higher than 0.8025 AUC  within 2 epochs with different settings of block sizes. The details are given in table \ref{tab:main_z}
\begin{table}[]
\centering
\caption{CriteoTB dataset: $1000\times$ reduced memory with \rmaz for varying $Z$. AUC was reached in $1.89$ epochs for all settings.}\label{tab:main_z}
\begin{tabular}{|l|l|}
\hline
Model($100$MB embedding) & $0.8025$ AUC reached? \\ \hline
\rma-$1$  & Yes                  \\ \hline
\rma-$8$  & Yes                  \\ \hline
\rma-$32$ & Yes                  \\ \hline
\rma-$128$ & Yes                  \\ \hline
\end{tabular}
\vspace{-0.5cm}
\end{table}
As we can see we can achieve the same target AUC, although with almost 2x time in terms of iterations. We experiment with different block sizes and see that we can achieve the required quality with different block sizes.
The results can be reproduced using our code. (see section \ref{sec:reproduce} in appendix for code details) 

\subsection{1000$\pmb{\times}$ Compression of Embedding Tables on Criteo Kaggle Dataset} \label{sec:kagacc}

For more comprehensive study on different state-of-the-art recommendation models, we use criteo kaggle dataset.

\textbf{Dataset:}
The Criteo Kaggle dataset (see section \ref{sec:reproduce} in appendix) has 13 integer features and 26 categorical features with 33.7M total categorical values. It is similar to CriteoTB dataset with lesser number of days and different sampling strategy. We split the data randomly into partitions 9:1, the smaller partition being used for testing. The training partition is further divided into partitions 8:2, the smaller partition being used for validation. We use early stopping based on validation AUC to choose the model. 

\textbf{Models}    
We use six different embedding based models from the literature: DLRM~\cite{DLRM19}, DCN~\cite{DCN17}, AutoInt~\cite{song2019autoint}, DeepFM~\cite{guo2017deepfm}, xDeepFM~\cite{lian2018xdeepfm}, and FiBiNET~\cite{huang2019fibinet}. The exact details of hyperparameters for the models and optimizer parameters, data split used for testing, and properties of the dataset used can be found in Appendix~\ref{app:main}. Specifically, we use embedding size of 16 for all the categorical values (around 33.7M). Hence, the original models have 540M parameters. We use Adam\cite{adam} for all models except DLRM which uses SGD as provided in original code. In this experiment, we set the compressed memory size to 540K parameters for \rmaz (i.e. $1000\times$ compression) 

\textbf{Results:}
Table \ref{tab:main} shows the results of AUC and along with the corresponding standard deviations for all the models in Table~\ref{tab:main}. The standard deviations of AUC of all settings are pretty similar and we exclude putting the results for other models (original and \rmaz for $Z > 1$) to save space.

We make the following observations from Table~\ref{tab:main}.
\begin{itemize}[nosep,leftmargin=*]
\item  Test AUC of \rmaz 1000 $\times$ compressed model is either better than original model (3/6 models) or similar (2/6 models). Only in case of XDeepFM , \rmaz performs worse than original model.

\item  The quality of model (i.e. AUC) reached is stable across different values of $Z$ for \rmaz.

\item As we can see, the number of epochs required to train the model of same quality is larger for \rmaz models as compared to original models. This is inline with our observation with CriteoTB dataset as well.
\end{itemize}
Our results can be reproduced using the DLRM code  for DLRM model and deep-torch code for other models( see section \ref{sec:reproduce} in appendix in appendix for code details)
\subsection{Inference Time for \rma-Z } \label{sec:inf}

\begin{table}[h]
\centering
\caption{Sample throughput: run with a batch size of 16384. The time includes the time taken to send the batch from RAM to GPU global memory and then the forward pass on the batch. The time also includes hash computation. There is 120\% increase in throughput by using \rma-$1$ which can be further improved using \rma-32 to 209\%. Original model is run on 4 QUADRO RTX 8000 GPUs while \rmaz models are run on a single GPU. All models have access to 120 CPUs. (CPUs are not involved in the measured computation though) }
\begin{tabular}{|c|c|c|}
\hline
Model                  & samples/second & Improvement \\ \hline
Original(100GB)        & 341454         & -           \\ \hline
\rma-1  & 755469         & 121\%       \\ \hline
\rma-2  & 865757         & 153\%       \\ \hline
\rma-8  & 913893         & 167\%       \\ \hline
\rma-32 & 920183         & 170\%       \\ \hline
\rma-128 & 1055470         & 209\%       \\ \hline
\end{tabular}

\label{tab:inferencetime}
\end{table}

With our proof-of-concept code (experimental and un-optimized), we measure the throughput of the samples during inference on CriteoTB dataset. We can see a phenomenal improvement in throughput during inference. While original 100GB model, run on 4 Quadro RTX-8000 (46GB) GPUs, can process around 341K samples per second, the \rmaz models which are only 100MB large, perform much faster. Using \rmaz we can process about $3.1\times$  samples. Specifically, we see 120\%($2.2\times$) improvement in throughput with $Z=1$. As expected, increasing value of $Z$ in \rmaz improves the throughput further upto 209\% ($3.1\times$) for \rma-128.

\subsection{Training Time for \rmazbold} \label{sec:train}
As noted in both the datasets, the running time of \rmaz models is slower w.r.t number of epochs required to reach the same quality. This can potentially be explained by the fact that having more parameters to tune can significantly speed up the learning. We see that with recommendation models with large embedding tables, we can achieve same quality with smaller compressed \rmaz given enough training time. A lot of research on recommendation models on click-through rate (CTR) data like Criteo, restrict themselves to 1 epoch of training. However, we want to stress that smaller models can potentially reach same quality and in these cases and it is just a matter of training more. Also, we believe that leveraging the smaller size of overall embeddings the training for these smaller \rmaz models can be faster than the original models w.r.t wall clock time. Our current timing experiments on proof-of-concept code ( un-optimized) does not show any improvements in training time. However, we plan to perform rigorous training time experiments in future with optimized code.

\vspace{-0.2cm}
\section{Conclusion}
\vspace{-0.1cm}

While industrial scale recommendation models are exploding due to large number of categorical features, \rmaa is a perfect alternative to embedding tables and enable training models of $1000 \times$ less memory to achieve same quality. \rmaa also shows clear inference throughput benefit and can potentially be trained much faster than original models. Also, training models with \rmaa is accessible to a average machine learning user who does not have access to high end hardware or engineering expertise required to train hundreds of TBs sized model. We believe DLRM with \rmaa will serve as a new baseline for compression and expedite the research in recommendation models. 

\bibliography{arxiv_main}
\bibliographystyle{unsrt}

\onecolumn
\section{Appendix}
\subsection{Reproducing results}\label{sec:reproduce}
We rely on the following repositories of code 

\begin{itemize}
    \item Official Model for CriteoTB : https://github.com/facebookresearch/dlrm/tree/6d75c84d834380a365e2f03d4838bee464157516
    \item DLRM Patch(to run dlrm model on kaggle and criteotb dataset) : https://github.com/apd10/dlrm
    \item robe-z code : https://github.com/apd10/universal\_memory\_allocation 
    \item kaggle challenge data : https://www.kaggle.com/c/criteo-display-ad-challenge
    \item deep-torch code (to run multiple state-of-the-art models on kaggle dataset) : https://github.com/apd10/criteo\_deepctr
\end{itemize}

\begin{itemize}
    \item Reproduce TB/kaggle results with DLRM model: 
        \begin{itemize}
            \item Install the robe-z code
            \item apply dlrm patch on dlrm original code 
            \item setup data for criteoTB/kaggle
            \item see the md file in dlrm file for commands
        \end{itemize}
    \item Reproduce kaggle results on other models: 
        \begin{itemize}
            \item Install the robe-z code
            \item use the deep-torch code
            \item run the criteo\_train file with appropriate config as provided in md file
        \end{itemize}
\end{itemize}

\subsection{\rmazbold}
    The parameter vector is constructed by flattening out the embedding table row-wise and concatenate all the embedding tables. let all the embeddings be of dimension $d$ and let the chunk size be $Z$. 
    
    The \rmaz is performed as follows
    \begin{itemize}
        \item split the parameter vector into chunks of $Z$. 
        \item hash each chunk to a particular  location in the array of size m 
        \item This chunk is added to the corresponding sub-array of the memory and in case we run outside of $m$ we cycle through to add at the beginning of the array. So the array we are sketching the parameter vector is actually a circular array
        \item each element is actually multiplied by the sign which is obtained by using another hash function g() and this is applied at the element level.
        We do not use the sign in our experiments. However, it can be used and greatly simplifies the theory.
    \end{itemize}
    
\subsection{Analysis 1: Analysis of quality of dimensionality reduction - feature hashing}

Let the parameter vector be in $\mathbb{R}^n$.  The projection maps this vector into $\mathbb{R}^m$.

Let $x \in \mathbb{R}^n$ and $y \in \mathbb{R}^n$
The estimator we want to analyse is that of inner product estimation . The estimator can be written in terms of the indicator functions $\mathds{1}(\cdot)$ as follows:

\begin{equation}
    \widehat{\ip{x}{y}} = \sum_{j = 1 }^m \left( \sum_{i=1}^n x_i g(i) \mathds{1}(h(i) == j) \right) \left( \sum_{i=1}^n y_i g(i) \mathds{1}(h(i) == j) \right).
\end{equation}

We can simplify the above indicator as

\begin{equation}
    \widehat{\ip{x}{y}} =  \sum_{i=1}^n \sum_{j=1}^n x_i y_j \mathds{1}(h(i) == h(j)),
\end{equation}
\begin{equation}
    \widehat{\ip{x}{y}} = \ip{x}{y} + \Sigma_{i \neq j}  x_i y_j \mathds{1}(h(i) {==} h(j)) g(i) g(j).
\end{equation}
Let $\mathcal{C}_i$ be the chunk-id that is assigned to to $i$, following the same notation used in Equations (\ref{eq:eq_rmax}).
Then, we know that $\mathds{1}(h(i) {==} h(j))$ is 0 if $\mathcal{C}_i {==} \mathcal{C}_j$.
Using this fact, we have
\begin{equation}
    \widehat{\ip{x}{y}} = \ip{x}{y} + \Sigma_{\mathcal{C}_i \neq \mathcal{C}_j}  x_i y_j \mathds{1}(h(i) {==} h(j)) g(i) g(j).
\end{equation}
We can easily see that this estimator of $\ip{x}{y}$ is unbiased.
Let us now look at

\begin{equation}
    \mathbb{E}(\widehat{\ip{x}{y}}) = \ip{x}{y}.
\end{equation}

The variance of the estimator can be computed as 
\begin{equation}
    \mathbb{V}(\widehat{\ip{x}{y}}) = E ((\widehat{\ip{x}{y}} - \ip{x}{y})^2) =  \mathbb{E}((\Sigma_{\mathcal{C}_i \neq \mathcal{C}_j}  x_i y_j \mathds{1}(h(i) {==} h(j)) g(i) g(j))^2),
\end{equation}
\begin{equation}
    \mathbb{V}(\widehat{\ip{x}{y}}) =  \mathbb{E}((\sum_{\mathcal{C}_i \neq \mathcal{C}_j, \mathcal{C}_{i'} \neq \mathcal{C}_{j'}}  x_i y_j \mathds{1}(h(i) {==} h(j)) g(i) g(j)) x_{i'} y_{j'} \mathds{1}(h(i') {==} h(j')) g(i') g(j')).
\end{equation}

The expected value of term in summation above is non-zero only if pairs are equal to eliminate the $g$s. As $i$ cannot be equal to $j$ , $i=j', i'=j$ or $i=i' , j'=j$. This implies that 

\begin{equation}
    \mathbb{V}(\widehat{\ip{x}{y}}) =  \mathbb{E}((\sum_{\mathcal{C}_i \neq \mathcal{C}_j}  x_i^2 y_j^2 \mathds{1}(h(i) {==} h(j))   + x_iy_ix_j y_j \mathds{1}(h(i) {==} h(j)).
\end{equation}

\textbf{Note that although there are some constraints that appear when we do chunk hashing. like if $\pmb{i}$ and $\pmb{j}$ collide then and $\pmb{i}$ and $\pmb{j}$ lie within the chunk, then $\pmb{i+1}$ and $\pmb{j+1}$ will also collide. But you can see that this relation does not really appear in the equation for variance. Maybe this appears in higher moments of the estimator.}

Using the fact that $\mathbb{E}(g(i)) = 0$, we can simplify the expression above as
\begin{equation}
    \mathbb{V}(\widehat{\ip{x}{y}}) = \frac{1}{m} (\Sigma_{\mathcal{C}_i \neq \mathcal{C}_j}  x_i^2 y_j^2 +  \Sigma_{\mathcal{C}_i \neq \mathcal{C}_j} x_i y_i x_j y_j). \label{eq:var}
\end{equation}
Note that when $Z=1$, the equation for variance is exactly the random projection that is used for "feature hashing" as proposed by \cite{featurehash}.

Let us denote the variance of inner product of vectors $x$ and $y$ projected from the dimension n to m while using chunk size of $Z$ to be $\mathbb{V}(x,y,Z,n,m)$. We will use this notation so that we are very precise in our statements.

Note that 
\begin{equation}
    \mathbb{V}(x,y,1,n,m) = \frac{1}{m} (\Sigma_{\mathcal{C}_i \neq \mathcal{C}_j}  x_i^2 y_j^2 +  \Sigma_{\mathcal{C}_i \neq \mathcal{C}_j} x_i y_i x_j y_j)  + \frac{1}{m} \sum_{c \in chunks}  (\Sigma_{i_c \neq j_c}  x_{i_C}^2 y_{j_c}^2 +  \Sigma_{i_c \neq j_c} x_{i_c} y_{i_c} x_{j_c} y_{j_c}),
\end{equation}
where $x_{i_c}$ is an element of sub-vector $x_c$, which refers to the chunk of the parameter vector.

\begin{equation}
    \mathbb{V}(x,y,1,n,m) = \mathbb{V}(x,y,Z,n,m)  +  \sum_{c \in chunks} \mathbb{V}(x_c, y_c, 1,Z, m )
\end{equation}
It is clear from the above equation that \rmaz has better variance w.r.t feature hashing .

\subsection{Effect of \rmazbold on inner product of embeddings of two values - i.e. parts of parameter vector that are identified as two separate embeddings. }

The previous analysis was the analysis of the sketching matrix and how good it is in preserving distances in a space. However, another important aspect of this projection - pertinent to the discussion of this paper is how does this projection of entire parameter vector affect the inter- embedding relation between embeddings of two different values. 

Consider how the parameter vector is constructed. We flatten out each embedding table row wise ( so each embedding is contiguous) and we concatenate all flattened embedding tables together to get a single parameter vector which is projected down.

There are three cases that we need to check. We will assume that either $Z$ divides $d$ or $d$ divides $Z$ (if $Z > d$ ) also, both Z and d divided n. Let the embeddings of two values under consideration be $x$ and $y$ 
\begin{itemize}
    \item $x$ and $y$ lie in same chunk ($Z> d$)
    \item $x$ and $y$ lie in different chunks $Z > d$
    \item $ Z < d$
\end{itemize}

\textbf{CASE: $\pmb{Z > d}$, $\pmb{x}$ and $\pmb{y}$ lie in the same chunk} $\pmb{\mathcal{C}}$

Let us first look at the product of two elements $x_1 y_1$.

\begin{equation}
    \hat{x_1} = x_1 + \sum_{i=1, i \notin \mathcal{C}}^n \theta_i g(x_1) g(i) \mathds{1}(h(i) == h(1))
\end{equation}

\begin{equation}
    \hat{y_1} = y_1 + \sum_{i=1, i \notin \mathcal{C}}^n \theta_i g(y_1) g(i) \mathds{1}(h(i) == h(y_1))
\end{equation}

\begin{equation}
\begin{split}
    &\hat{x_1 y_1} = 
    x_1 y_1 + \\ 
    &x_1 (\sum_{i=1, i \notin \mathcal{C}}^n \theta_i g(y_1) g(i) \mathds{1}(h(i) == h(y_1))) + y_1 (\sum_{i=1, i \notin \mathcal{C}}^n \theta_i g(x_1) g(i) \mathds{1}(h(i) == h(1))) + \\
    &\sum_{i=1, \notin \mathcal{C}, j=1, j \notin \mathcal{C}} \theta_i, \theta_j g(x_1) g(y_1) g(i) g(j) \mathds{1}(h(i) = h(x_1)) \mathds{1}(h(j) == 
    h(y_1))
\end{split}
\end{equation}

\begin{equation}
    \mathbb{E}(\widehat{x_1 y_1}) = xy
\end{equation}
Hence,
\begin{equation}
    \mathbb{E}(\widehat{\ip{x}{y}}) = \ip{x}{y}
\end{equation}

\textbf{CASE : $\pmb{Z > d}$ or $\pmb{Z \leq d}$, $\pmb{x}$ and $\pmb{y}$ lie in different chunks $\pmb{\mathcal{C}_1}$ and $\pmb{\mathcal{C}_2}$}

Let us first look at the product of two elements  $x_1 y_1$.

\begin{equation}
    \hat{x_1} = x_1 + \sum_{i=1, i \notin \mathcal{C}_1}^n \theta_i g(x_1) g(i) \mathds{1}(h(i) == h(1))
\end{equation}

\begin{equation}
    \hat{y_1} = y_1 + \sum_{i=1, i \notin \mathcal{C}_2}^n \theta_i g(y_1) g(i) \mathds{1}(h(i) == h(y_1))
\end{equation}

\begin{equation}
\begin{split}
    &\hat{x_1 y_1} = 
    x_1 y_1 + \\ 
    &x_1 (\sum_{i=1, i \notin \mathcal{C}_2}^n \theta_i g(y_1) g(i) \mathds{1}(h(i) == h(y_1))) + y_1 (\sum_{i=1, i \notin \mathcal{C}_1}^n \theta_i g(x_1) g(i) \mathds{1}(h(i) == h(1))) + \\
    &\sum_{i=1, \notin \mathcal{C}_1, j=1, j \notin \mathcal{C}_2} \theta_i, \theta_j g(x_1) g(y_1) g(i) g(j) \mathds{1}(h(i) = h(x_1)) \mathds{1}(h(j) == 
    h(y_1))
\end{split}
\end{equation}

\begin{equation}
    \mathbb{E}(\widehat{x_1 y_1}) = xy ( 1  +  \frac{1}{m})
\end{equation}
Hence,
\begin{equation}
    \mathbb{E}(\widehat{\ip{x}{y}}) = \ip{x}{y} ( 1 + \frac{1}{m})
\end{equation}

\textbf{Variance:}

We will analyse the variance for specific case of $Z > E=d$ and $x$ and $y$ values have embeddings in different blocks. Other cases can be computed similarly as

\begin{equation}
\begin{split}
     \widehat{\ip{\vec{\theta_x}}{\vec{\theta_y}}} = 
    \sum_{j=1}^d & \left( \sum_{i=1}^n \theta_i g(idx_x + j) g(i) \mathds{1}(h(i) == h(idx_x + j)) \right) \\ 
    & \quad \left( \sum_{i=1}^n \theta_i g(idx_y  + j) g(i) \mathds{1}(h(i) == h(idx_y + j)) \right)
    \end{split}
\end{equation}

\begin{equation}
\begin{split}
     & (\widehat{\ip{\vec{\theta_x}}{\vec{\theta_y}}})^2 = 
    \sum_{j_1 = 1}^d \sum_{j_2 = 1}^d \sum_{i_1 = 1}^n    \sum_{i_2 = 1}^n \sum_{i_3 = 1}^n \sum_{i_4 = 1}^n \\
    & \quad \quad \theta_{i_1} \theta_{i_2}\theta_{i_3} \theta_{i_4} \\
    & \quad \quad g(i_1) g(i_2) g(i_3) g(i_4)\\
    & \quad \quad g(idx_x +j_1) g(idx_y +j_1) g(idx_x +j_2) g(idx_y +j_2) \\
    & \quad \quad \mathds{1}(h(i_1) {=} h(idx_x + j_1)) \mathds{1}(h(i_2) {=} h(idx_y + j_1)) \mathds{1}(h(i_3) {=} h(idx_x + j_2)) \mathds{1}(h(i_4) {=} h(idx_y + j_2))
\end{split}
\end{equation}

separating cases when  ($i_1 = idx_x +j_1$, $i_2 = idx_y +j_1$ , $i_3 = idx_x + j_2$, $i_4= idx_y + j_2$) and others
\begin{equation}
\begin{split}
     & (\widehat{\ip{\vec{\theta_x}}{\vec{\theta_y}}})^2 = \sum_{k_1 = 1}^d \sum_{k_2 = 1}^d x_{k_1} y_{k_1} x_{k_2} y_{k_2} + \\
    & \sum_{j_1 = 1}^d \sum_{j_2 = 1}^d \sum_{i_1 = 1, i_1 \neq idx_x + j_1}^n    \sum_{i_2 = 1, i_1 \neq idx_y + j_1}^n \sum_{i_3 = 1, i_3 \neq idx_x + j_2}^n \sum_{i_4 = 1, i_4 \neq idx_y + j_2}^n \\
    & \quad \quad \theta_{i_1} \theta_{i_2}\theta_{i_3} \theta_{i_4} \\
    & \quad \quad g(i_1) g(i_2) g(i_3) g(i_4)\\
    & \quad \quad g(idx_x +j_1) g(idx_y +j_1) g(idx_x +j_2) g(idx_y +j_2) \\
    & \quad \quad \mathds{1}(h(i_1) {=} h(idx_x + j_1)) \mathds{1}(h(i_2) {=} h(idx_y + j_1)) \mathds{1}(h(i_3) {=} h(idx_x + j_2)) \mathds{1}(h(i_4) {=} h(idx_y + j_2))
\end{split}
\end{equation}

\begin{equation}
\begin{split}
     & \mathbb{V}({\ip{\vec{\theta_x}}{\vec{\theta_y}}}) = \\
    &\mathbb{E}( \sum_{j_1 = 1}^d \sum_{j_2 = 1}^d \sum_{i_1 = 1, i_1 \neq idx_x + j_1}^n    \sum_{i_2 = 1, i_1 \neq idx_y + j_1}^n \sum_{i_3 = 1, i_3 \neq idx_x + j_2}^n \sum_{i_4 = 1, i_4 \neq idx_y + j_2}^n \\
    & \quad \quad \theta_{i_1} \theta_{i_2}\theta_{i_3} \theta_{i_4} \\
    & \quad \quad g(i_1) g(i_2) g(i_3) g(i_4)\\
    & \quad \quad g(idx_x +j_1) g(idx_y +j_1) g(idx_x +j_2) g(idx_y +j_2) \\
    & \quad \quad \mathds{1}(h(i_1) {=} h(idx_x + j_1)) \mathds{1}(h(i_2) {=} h(idx_y + j_1)) \mathds{1}(h(i_3) {=} h(idx_x + j_2)) \mathds{1}(h(i_4) {=} h(idx_y + j_2)) )
\end{split}
\end{equation}

We will analyse the case of embeddings of $x$ and $y$ lie in separate blocks and $Z > d$.

\begin{table}[h]
\resizebox{\textwidth}{!}{
\begin{tabular}{|l|l|l|l|l|}
\hline
        &                                                 & term                                        & factor                 & comment                                                                                                                                                                                                                                                                                                                                                                                                                                                                                            \\ \hline
j1 = j2 & i1=i2=i3=i4                                     & t\textasciicircum{}4                        & 1/m\textasciicircum{}2 & \multirow{4}{*}{\begin{tabular}[c]{@{}l@{}}The value of i\_1, i\_2, i\_3, i\_4 should be equal\\ in pairs or all of them. Also, they should not match their\\ own elements. ( i.e. i\_1 != x+j1, etc)\\ t = typical element of theta\end{tabular}}                                                                                                                                                                                                                                                 \\ \cline{1-4}
        & i1 = i2 != i3=i4                                & t1\textasciicircum{}2 t2\textasciicircum{}2 & 1/m\textasciicircum{}4 &                                                                                                                                                                                                                                                                                                                                                                                                                                                                                                    \\ \cline{1-4}
        & i1 = i3 != i2 = i4                              & t1\textasciicircum{}2 t2\textasciicircum{}2 & 1/m\textasciicircum{}2 &                                                                                                                                                                                                                                                                                                                                                                                                                                                                                                    \\ \cline{1-4}
        & i1 = i4 != i2 =i3                               & t1\textasciicircum{}2 t2\textasciicircum{}2 & 1/m\textasciicircum{}4 &                                                                                                                                                                                                                                                                                                                                                                                                                                                                                                    \\ \hline
j1!=j2  & i1 = y + j1, i3 = y + j2 , i2 = x + j1, i4=x+j2 & x1y1x2y2                                    & 1/m\textasciicircum{}2 & \multirow{12}{*}{\begin{tabular}[c]{@{}l@{}}The value of i\_1, i\_2, i\_3, i\_4 should be equal to \\ the  4 distinct elements ( 2 of x and 2 of y) without\\ matching their own elements.\\ Also, as the indicators should be non-zero\\ So,\\ i\_1 cannot be x+j1 also it cannot be x+j2 as the value of \\ indicator in that case will always be 0. Since Z \textgreater d h(i\_1=x+j2) \\ cannot be equal to h(x+j1)\\ Hence i\_1 can be equal to y+j1 or y+j2\\ x,y are vectors\end{tabular}} \\ \cline{1-4}
        & i1 = y + j1, i3 = y + j2 , i2 = x + j2, i4=x+j1 & x1y1x2y2                                    & 1/m\textasciicircum{}3 &                                                                                                                                                                                                                                                                                                                                                                                                                                                                                                    \\ \cline{1-4}
        & i1 = y + j2, i3 = y + j1 , i2 = x + j1, i4=x+j2 & x1y1x2y2                                    & 1/m\textasciicircum{}3 &                                                                                                                                                                                                                                                                                                                                                                                                                                                                                                    \\ \cline{1-4}
        & i1 = y + j2, i3 = y + j1 , i2 = x + j2, i4=x+j1 & x1y1x2y2                                    & 1/m\textasciicircum{}2 &                                                                                                                                                                                                                                                                                                                                                                                                                                                                                                    \\ \cline{1-4}
        &                                                 &                                             &                        &                                                                                                                                                                                                                                                                                                                                                                                                                                                                                                    \\ \cline{1-4}
        &                                                 &                                             &                        &                                                                                                                                                                                                                                                                                                                                                                                                                                                                                                    \\ \cline{1-4}
        &                                                 &                                             &                        &                                                                                                                                                                                                                                                                                                                                                                                                                                                                                                    \\ \cline{1-4}
        &                                                 &                                             &                        &                                                                                                                                                                                                                                                                                                                                                                                                                                                                                                    \\ \cline{1-4}
        &                                                 &                                             &                        &                                                                                                                                                                                                                                                                                                                                                                                                                                                                                                    \\ \cline{1-4}
        &                                                 &                                             &                        &                                                                                                                                                                                                                                                                                                                                                                                                                                                                                                    \\ \cline{1-4}
        &                                                 &                                             &                        &                                                                                                                                                                                                                                                                                                                                                                                                                                                                                                    \\ \cline{1-4}
        &                                                 &                                             &                        &                                                                                                                                                                                                                                                                                                                                                                                                                                                                                                    \\ \hline
\end{tabular}
}
\caption{For x and y lie in different blocks and Z \textgreater d . so each embedding lies completely inside the block.}
\label{tab:my-table}
\end{table}
For convenience we are using $x = \vec{\theta_x}$ and similarity for $y$
\begin{equation}
\begin{split}
     \mathbb{V}({\ip{\vec{\theta_x}}{\vec{\theta_y}}})  \leq  d \frac{1}{m^2} \sum_{i=1}^n \theta_i^4 
     +2 d \frac{1}{m^4} \sum_{i \neq j} \theta_i^2 \theta_j^2 +
    d \frac{1}{m^2} \sum_{i \neq j} \theta_i^2 \theta_j^2 +  2(\frac{1}{m^2} + \frac{1}{m^3})\sum_{i\neq j = 1}^d x_i y_i x_j y_j)
\end{split}
\end{equation}

We use less than as not all elements from $\theta$ are present in actual summation.

\begin{equation}
\begin{split}
     \mathbb{V}({\ip{\vec{\theta_x}}{\vec{\theta_y}}})  \leq  \mathbb{O}\left( \frac{d}{m^2} ||\theta||^4 + \frac{1}{m^2} (\sum_{i\neq j = 1}^d x_i y_i x_j y_j) \right)
\end{split}
\end{equation}

\subsection{Criteo Kaggle Experiment} \label{app:main}
The experimental settings are described in detail below

\textbf{Dataset:} We choose the Criteo Kaggle dataset in order to demonstrate the compressive power of UMA. The original dataset of Criteo has 13 integer features and 26 categorical features. The counts of the feature values in total is around 33M. the breakup in each of the feature is as follows.
\begin{verbatim}
counts:
[1460, 583, 10131227, 2202608, 305, 24, 12517, 633, 3, 93145, 5683, 8351593, 
 3194, 27, 14992, 5461306, 10, 5652, 2173, 4, 7046547, 18, 15, 286181, 105,
 142572]
\end{verbatim}
We do not perform any rare feature filtering, which reduces the number of categorical values as is done in papers reporting SOTA values for the Criteo dataset. We want to demonstrate an algorithm that can deal with large embedding tables (e.g., terabytes-size of embedding tables as observed in industries) and choose the entire feature set of the Criteo dataset to make a reasonable dataset. Also, this is standard practice in papers which deal with ``compression'' or efficient embedding tables research \cite{MDTrick19, QuoRemTrick19}.

\textbf{Hyperparameters chosen for different models.}
The hyperparameters for running different models were chosen as mentioned in the respective original papers. We did not do hyperparameter tuning for UMA as the result sufficiently supports our hypothesis that these memory allocation approaches are valuable to efficiently learn compressed embedding tables. We fix the batch size of each of training to 2048 and cutoff the training at 300K iterations for all models except DLRM which is run till 540K (due to SGD). All the models have embedding size = 16.
\begin{table}[h]
\centering
\caption{Hyperparameters of different model chosen as per the specification in their papers. The code used for running DLRM model is : https://github.com/facebookresearch/dlrm . For other models, the code used is https://github.com/shenweichen/DeepCTR-Torch.}
\resizebox{\textwidth}{!}{
\begin{tabular}{|c|c|c|c|c|c|}
\hline
        & architecture                                                                                       & dropout & l2\_regularization & optimizer & learning rate                                               \\ \hline
DLRM    & \begin{tabular}[c]{@{}c@{}}bot:   13-512-256-64-16\\ top:   512-256-1\end{tabular}                 & 0       & 0                  & SGD       & \begin{tabular}[c]{@{}c@{}} 1.0 \end{tabular} \\ \hline
DCN     & 1024-1024-1024-1024                                                                                & 0       & 0                  & ADAM      & 0.001                                                       \\ \hline
AutoInt & \begin{tabular}[c]{@{}c@{}}400-400-400\\ attention\_embedding\_size:\\ 32\end{tabular}             & 0       & 0                  & ADAM      & 0.001                                                       \\ \hline
DeepFM  & 400-400-400                                                                                        & 0.5     & 0                  & ADAM      & 0.001                                                       \\ \hline
XDeepFM & \begin{tabular}[c]{@{}c@{}}dnn:\\ 400-400-400\\ cross interaction: \\     200-200-200\end{tabular} & 0.5     & 0.0001             & ADAM      & 0.001                                                       \\ \hline
FiBiNET & 400-400-400                                                                                        & 0.5     & 0.0001             & ADAM      & 0.001                                                       \\ \hline
\end{tabular}
}
\label{tab:hyperparams}
\end{table}

\textbf{Train/Test data split.}
We use the following way to split the data into random training and testing data for all models except DLRM. For DLRM, the code provided has their own data loader and we do not make any changes there.
\begin{verbatim}
from sklearn.model_selection import train_test_split
train, test = train_test_split(data, test_size=0.1, random_state=2020)
\end{verbatim}

\textbf{Choosing models while training.}
We observe that most models (especially the original models) start to overfit after some iterations. Hence, we use early stopping based on validation AUC to select the final model.

\subsubsection{FAQ on Kaggle Experiment:}
\begin{enumerate}
    \item \textbf{Why do we not perform rare feature filtering?}
    We do not perform any sort of rare feature filtering which reduces the number of categorical values as is done in papers reporting SOTA values for criteo kaggle dataset. It is important to note that such tricks that are optimized on smaller public benchmarks do not help in the real production scale datasets \cite{gupta2020architectural} and hence are avoided. We want to demonstrate an algorithm that can deal with large embedding tables (e.g., terabytes-size of embedding tables as observed in industries) and choose the entire feature set of the Criteo dataet to make a reasonable dataset. Also, this is a standard in papers which deal with ``compression'' or efficient embedding tables research \cite{MDTrick19, QuoRemTrick19}. 
    
    \item \textbf{Why do we not achieve the SOTA results (as reported on original papers)?}
    Reproducing results in the original papers would require access to their codes and dataset preprocessing if any and dataset split they use to create training and testing datasets. Unfortunately, most papers do not have their own codes public and do not specify the random seeds used to split the data. In our experiments, we use the random seed = 2020 for all the models except DLRM which has its own random splitter in the code provided. Apart from this, we also do not perform rare feature filtering which might affect the results. However, our experiments on some models with rare feature filtering showed that it does not help with performance of original model.
    \item \textbf{Why we use fixed embedding size for models like DCN, which tell us to use custom embedding sizes?} While it is true that DCN model gives a custom way to choose the size of the embedding based on the number of values in that category, using the formula leads to very large memory tables for the Criteo dataset using full features (no rare feature filtering). Hence, it is not possible to use the custom formula in our case. We uniformly set the embedding size to 16 across the different models.
\end{enumerate}

\end{document}